\documentstyle[11pt,paspconf]{article}

\begin{document}
\newcommand{\chvcs}{{C\,{\sc iv}}-HVCs}
\newcommand{\degr}{$^{\circ}$}
\newcommand{\et}{{\it et al.~}}
\newcommand{\hhvcs}{{H\,{\sc i}}-HVCs}
\newcommand{\kms}{\,km\,s$^{-1}$}     
\newcommand{\lam}{$\lambda$}
\newcommand{\lya}{Ly$\alpha$\ }
\newcommand{\subsun}{\mbox{$_{\odot}$}}
\newcommand{\twid}{\,$\sim$\,}
\newcommand{\vlsr}{V$_{\sc LSR}$}
\newcommand{\phn}{\phantom{0}}

\def\hi{H\,{\sc i}}
\def\cii{C\,{\sc ii}}
\def\civ{C\,{\sc iv}}
\def\nv{N\,{\sc v}}
\def\oi{O\,{\sc i}}
\def\ovi{O\,{\sc vi}}
\def\siii{Si\,{\sc ii}}
\def\siiv{Si\,{\sc iv}}

\hyphenation{additional}
\hyphenation{Telescope}
\hyphenation{under-stand}
\hyphenation{obtained}
\hyphenation{toward}
\hyphenation{involving}
\hyphenation{viability}
\hyphenation{ratios}
\hyphenation{radi-ation}
\hyphenation{obser-va-tions}
\hyphenation{direc-tion}
\hyphenation{extra-galactic}
\hyphenation{cosmological}

\title{Highly Ionized High Velocity Clouds}

\author{Kenneth R. Sembach}
\affil{Department of Physics \& Astronomy, Johns Hopkins University,
Baltimore, MD 21218, U.S.A}

\author{Blair D. Savage}
\affil{Department of Astronomy, University of Wisconsin, Madison, WI 53706}

\author{Limin Lu}
\affil{Lucent Technologies, Naperville, IL  60566}

\author{Edward M. Murphy}
\affil{Department of Physics \& Astronomy, Johns Hopkins University,
Baltimore, MD 21218, U.S.A}

\begin{abstract}
We have recently used the {\it Hubble Space Telescope} to study
a pair of high velocity clouds in the direction of 
Mrk\,509 that have unusual ionization properties. They exhibit strong \civ\
absorption with little or no low ion absorption or \hi\ 21\,cm emission. 
As the closest known analog to the outer diffuse halos of damped Ly$\alpha$
absorbers and the low N(\hi) metal line absorption systems
seen in the spectra of high redshift quasars, these ``\chvcs'' may shed
new light on the origins of some HVCs, as well as present  
opportunities for comparing absorption due to  intergalactic gas in the
local universe
with absorption in moderate-high redshift gas clouds.
The \chvcs\ have ionization properties consistent with photoionization by 
extragalactic
background radiation and a location within the Local Group.
The presence of weak \hhvcs\ 
detected through 21\,cm emission within 2$\deg$ of the  
sight line suggests that the \chvcs\ trace extended, ionized,
low density regions of the \hhvcs.  
In this article we summarize the results of our study of the \chvcs\
and suggest additional observations that would
test the hypothesis of
an intergalactic location for the clouds.
\end{abstract}

\keywords{ISM: clouds--- ISM: atoms --- Galaxy: halo --- ultraviolet: ISM}

\section{Introduction}

Investigations of galactic chemical evolution and galaxy formation at
moderate to high redshift have progressed rapidly in recent years due to 
spectroscopic observations from the ground and imaging campaigns
with the {\it Hubble Space Telescope} (HST).  To tie these
studies into the present-day epoch, it is necessary to understand the
processes that govern galactic evolution within the local universe.  
Determining the properties of high velocity clouds (HVCs), which typically 
have velocities $|$\vlsr$|$\,$>$\,100 \kms, may prove to be 
an important step in revealing the processes that distribute 
and ionize gases in the halos of the Milky Way and other galaxies. 
These clouds can be studied spectroscopically without confusion from lower 
velocity absorption if suitable background sources can be found.

 There 
have been many ideas proposed to explain the distribution of HVCs on the 
sky and their kinematics, including supernova-driven ``Galactic fountains''
(Shapiro \& Field 1976; Bregman 1980), ram pressure or 
tidal stripping of material from the Magellanic Clouds (Moore \& Davis 1994; 
Lin \et 1995), infalling gas from outside the Milky Way (Oort 1970; 
Mirabel \& 
Morras 1984), and identification of HVCs as intergalactic clouds within the 
Local Group (Blitz \et 1998).  However, 
distinguishing between all of these possible 
scenarios has not been easy, despite many years of study 
(see Wakker \& van~Woerden 1997 for a review of HVC properties and 
possible origins).

Traditionally, HVCs have been studied through their \hi\ 21\,cm emission,
but it is now possible to make fundamental advances in understanding HVCs by
examining their absorption properties in detail with 
the HST.  We have obtained a full 
suite of {\it Goddard High
Resolution Spectrograph} (GHRS) absorption line 
observations to study the neutral and ionized gases in two HVCs 
in the direction of Mrk\,509 ($l\,=\,36.0\deg, b\,=\,-29.9\deg$).  
The clouds exhibit strong \civ\ 
absorption with little or no corresponding low ion (\cii, \siii) absorption 
or \hi\ 21\,cm emission down to a detection threshold of 
log\,N(\hi) $\approx$ 17.7 (see Figure~1).  
These ``\chvcs'' have ionization 
properties that 
are very different from those of gases in the Galactic disk and low halo
(Sembach \et 1995, 1999).  Figure~1
shows that the low velocity ($|$\vlsr$|$\,$<$\,100 \kms) gas in the 
Galactic disk and halo along the Mrk\,509 sight line
is characterized by both high ionization  and strong low 
ionization absorption.
This is typical for diffuse interstellar gases in the Milky Way
(Sembach \& Savage 1992),
but is very different than the C\,{\sc iv}-HVC absorption signature at
$|$\vlsr$|$\,$\approx$\,$-$340 to $-$170 \kms, where the dominant absorbing 
ion is clearly \civ.  
To a large degree, the \chvcs\ resemble the low column density 
(log\,N(\hi)\,$<$\,17) QSO metal line absorption systems, in which one often 
sees strong \civ\ (and sometimes \siiv) absorption but little low ionization 
absorption (Steidel 1990; Songaila \& Cowie 1996).

\section{Properties of the \chvcs}
Using the photoionization code CLOUDY (Ferland 1996), we modeled the \chvcs\ 
toward Mrk\,509 as slabs of gas bathed in 
extragalactic background radiation 
with a QSO spectral energy distribution  and a mean intensity at the 
Lyman limit J$_{\nu}$(LL) = 1$\times$10$^{-23}$ erg cm$^{-2}$ s$^{-1}$ 
Hz$^{-1}$ sr$^{-1}$ (Haardt \& Madau 1996).  
We show a sample model from Sembach \et (1999) 
in Figure~2.  In this model, the column densities are satisfied for 
a narrow range of ionization parameters, $\Gamma$\,=\,n$_\gamma$/n$_H$.
The combination of $\Gamma$ constrained by the ionic ratios and an 
assumption about J$_{\nu}$(LL), which is proportional to n$_\gamma$, 
sets the density of the model cloud.  The size, D\,=\,N(H)/n$_H$,
is also determined
since the observed metal line column densities yield N(H) for a given
metallicity.

\begin{table}[!h]
\caption{Properties of the \chvcs\ Toward Mrk\,509$^a$}
\begin{center}
\small
\begin{tabular}{lccccccc}
\tableline
\tableline
\vlsr\ & log\,N(H)  & log\,N(\hi) & n$_H$/n$_{HI}$ & log\,n$_H$  & T & P/k  & Diam. \\
(\kms)& (cm$^{-2}$) & (cm$^{-2}$) && (cm$^{-3}$) & (K) & (cm$^{-3}$\,K) & (kpc) \\
\tableline
$-$283  & 19.05 & 16.30 &  560 & $-$3.82 & 14,800 & $\sim$5 & 28 \\
$-$228  & 18.16 & 14.70 & 2900 & $-$4.48 & 19,200 & $\sim$1 & 16 \\
\tableline
\end{tabular}
\end{center}
\footnotesize
\tablenotetext{a}{ Values in this table are appropriate for ionization of a 
uniform slab of
material bathed in extragalactic background radiation at $z$\,=\,0.  The
calculations are for a model with a metallicity of 1/3 solar
and no dust.  Lower metallicity models yield approximately the same results,
except that the cloud sizes are proportionally larger.  For example,
the cloud diameters for [Z/H]\,=\,$-$1 are 108 and 62 kpc for the $-$283 and
$-$228 \kms\ clouds, respectively.  Note the very low
thermal pressures, which are much smaller than values for typical diffuse gas 
in the Milky Way where P/k\,$\sim$\,4$\times$10$^3$~cm$^{-3}$~K.}
\end{table}

We compared the observations to
predictions of models involving collisional ionization or
photoionization by starlight and found that   
the ionic ratios are most consistent with
 photoionization by extragalactic background radiation. 
If the gas is photoionized by the extragalactic background or a 
combination of ultraviolet starlight and the extragalactic background, the 
clouds must be low density (n$_H$\,$\sim$\,10$^{-4}$~cm$^{-3}$), 
large (greater than several kiloparsecs), and 
mostly ionized (n$_{HI}$/n$_H$\,$\sim$\,10$^{-3}$) regions 
located well beyond the neutral gas layer of the Galaxy.  
If the clouds are intergalactic in nature, their metallicities could be
 [Z/H]\,$\sim$\,$-$1 or 
lower.  Table~1 contains a summary of the C\,{\sc iv}-HVC properties,
assuming [Z/H]\,=\,$-$0.5.  

The low inferred pressures for the \chvcs,
P/k\,$\sim$\,1--5 cm$^{-3}$~K,
are strongly suggestive of an intergalactic origin.
The  pressures 
are 2--3 orders of magnitude smaller than those predicted for 
multi-phase models of the Galactic halo at $|$z$|$\,$<$\,10 kpc (e.g., 
Wolfire \et 1995).

\begin{figure}[!ht]
\includegraphics{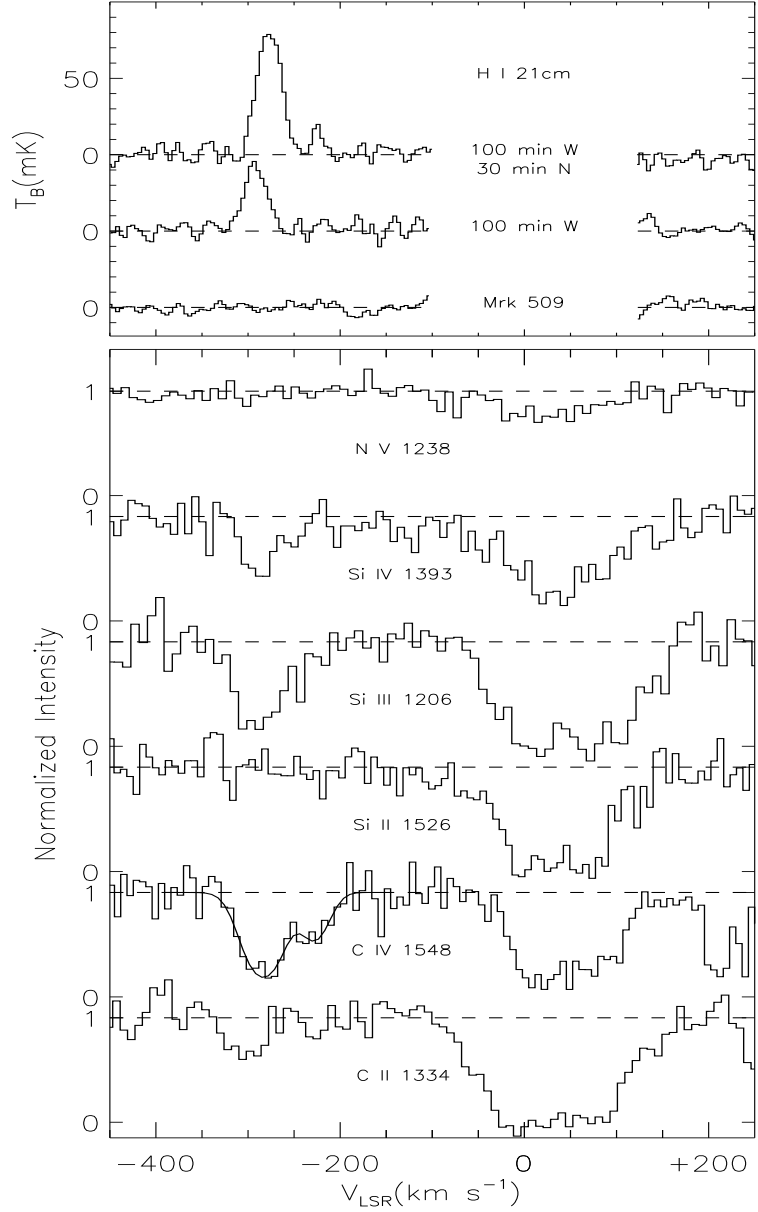}
\vspace{6.4in}
\caption{\footnotesize {\it Top:} Brightness temperature versus LSR 
velocity for \hi\ 21\,cm emission toward Mrk\,509 and two nearby 
positions on the sky.  Note
the absence of \hi\ emission at high velocities directly along the sight line.
{\it Bottom:} Continuum normalized intensity versus LSR velocity for the 
\nv\,$\lambda$1238, \siiv\,$\lambda$1393, Si\,{\sc iii}\,$\lambda$1206,
\siii\,$\lambda$1526, \civ\,$\lambda$1548, and \cii\,$\lambda$1334 lines 
observed by Sembach \et
(1999) with the GHRS G160M grating (R\,$\sim$\,20,000). Note the presence of 
strong \civ\ absorption at velocities between $-$340 and $-$170 \kms.  The
absorption in these
``\chvcs'' is unlike the absorption due to lower velocity gas 
tracing the Milky Way disk and halo along the sight line.  The \civ-HVC 
absorption probably arises in Local Group gas subjected to ionizing 
extragalactic background radiation.}
\end{figure}

\begin{figure}[!h]
\includegraphics{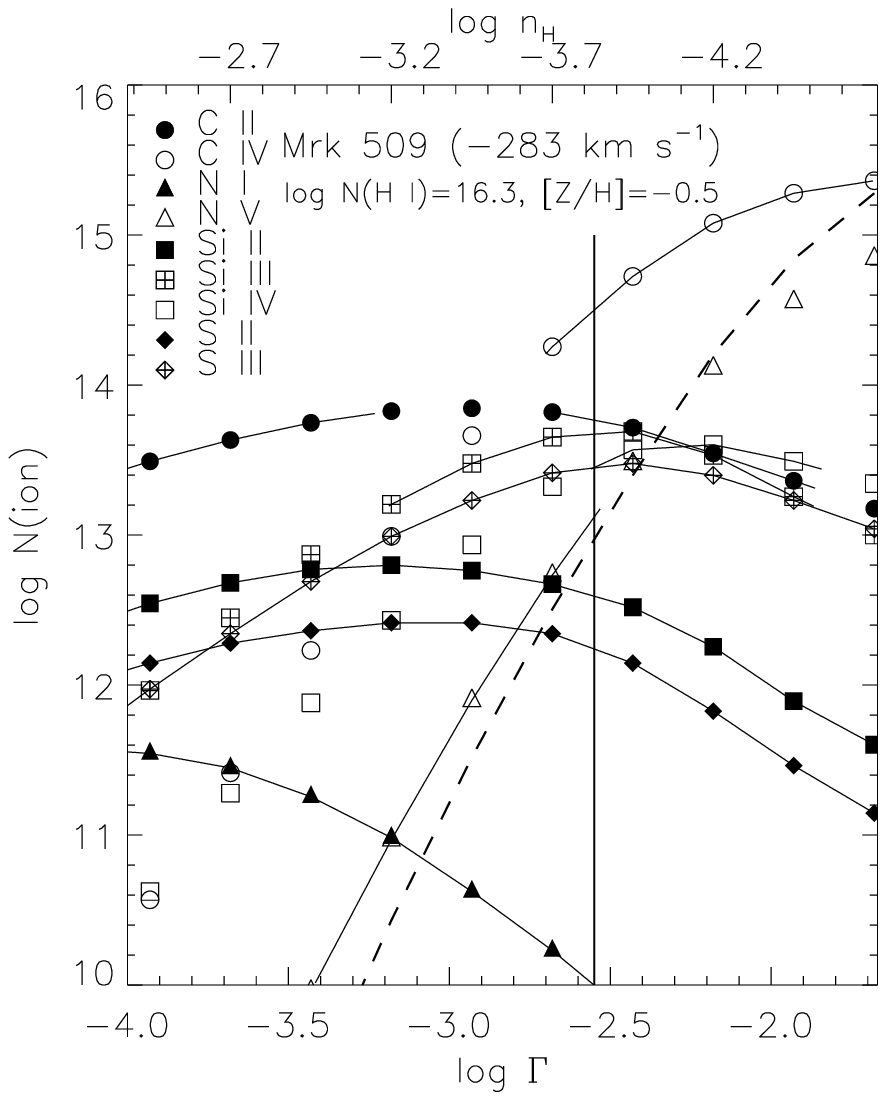}
\vspace{3.2in}
\caption{\footnotesize Photoionization model calculation for the
$-$283~km~s$^{-1}$
C\,{\sc iv}-HVC toward Mrk\,509.  The points indicate the model 
predictions of the 
ionic column densities as a function of ionization parameter, $\Gamma$ 
(= n$_\gamma$/n$_H$). Solid lines indicate the portions of the ion curves 
that satisfy the observational constraints.  The vertical line indicates
the value of log\,$\Gamma$\,=\,$-$2.55 (or log\,n$_H$\,=\,$-$3.82) that 
allows a simultaneous fit to 
all the observable constraints.  The dashed line is a prediction for
O~VI.  The physical parameters of the model are 
listed in Table~1.}
\end{figure}

\section{The Origin of HVCs and Relevance to Galaxy Evolution}
Blitz \et (1998) have recently proposed the intriguing hypothesis 
that some of the observed
\hhvcs\ in the sky are large (diam\,$\approx$\,25~kpc) dark-matter-dominated 
clouds 
located within the Local Group.  Their argument for an extragalactic 
location for the clouds is based upon the good agreement between the 
velocity centroid of the cloud ensemble considered and that of the 
Local Group (see Blitz, this volume).
They suggest that the Milky Way has accreted, and continues to accrete,
such clouds.  Such an accretion could have profound implications for the
physical state of the Galactic halo and the chemical evolution of the 
Galaxy, since the infalling gas clouds in this
model typically have masses of $\sim$\,10$^{8}$ M$_\odot$. 
The kinematics and ionization of the \chvcs\ toward Mrk\,509 are consistent 
with the Blitz \et (1998) model and suggest that an intergalactic 
origin for some HVCs may be possible. However, confirmation of an ensemble of
such clouds will require further observations and tests, such as direct
measurements of metallicities in the \hhvcs.

The properties of the \chvcs\ toward Mrk\,509 are similar to
the properties found for high redshift ($z$\,$\sim$\,3) Ly$\alpha$ clouds
with 10$^{14}$\,$<$\,N(\hi)\,$<$\,10$^{17}$ cm$^{-2}$ and 
log\,$\Gamma$\,$\approx$\,$-$2.5 to $-$1.9
(see Songaila \& Cowie 1996).  In recent
cosmological simulations of large-scale structure formation 
involving gas hydrodynamics, Ly$\alpha$ clouds at 
2\,$<z<$\,4 with 10$^{14}$\,$<$\,N(\hi)\,$<$\,10$^{17}$~cm$^{-2}$ trace the 
diffuse gas in the filamentary and sheet-like structures that surround and 
connect galaxies (Zhang \et 1995; Petitjean \et 1995;  
Hernquist \et 1996; Miralda-Escud\'e \et 1996).
The simulations indicate that the gas in the sheets and 
filaments gradually become denser concentrations that eventually form galaxies.
Some of this gas may have survived to 
the present day.  The \chvcs\ and some of the \hhvcs\ may be manifestations 
of such filamentary structures within the nearby universe.
Additional observations of \chvcs\ along other sight lines would 
provide important insight into the viability of this hypothesis.

The origins of HVCs have been debated for decades.  The \chvcs\ provide
fundamentally new, and challenging, information that has not been available
previously.  While the \chvcs\ trace mainly ionized gas, their study
has applications for understanding neutral HVCs as well.
We detect weak \hi\ emission within 2$\deg$ of the Mrk\,509
sight line.  The physical connection between the \chvcs\ and the \hhvcs\ 
is still uncertain, though the relative proximity of the \hhvcs\ and the 
velocity similarities of the \hhvcs\ to the \chvcs\ 
along the sight line suggests that the two types of HVCs are related.  The 
simplest relationship would be one in which the \chvcs\ 
trace the extended, ionized, low density regions of the \hhvcs.

\section{Where Do We Go From Here?}
There are several observations that could provide further insight into the
nature of the \chvcs.  

\subsection{H$\alpha$ Emission}
Considerations of different scenarios for the production of H$\alpha$
emission (Ferrara \& Field 1994; Bland-Hawthorn \et 1995; 
Bland-Hawthorn 1997) illustrate that it can be a powerful discriminator 
between an extragalactic and a Galactic location for the \chvcs.
The \chvcs\ should have very low levels
of H$\alpha$ emission if the clouds are extragalactic entities photoionized
by extragalactic background radiation.   
If, however, the clouds are 
local (i.e., in the Milky Way halo), H$\alpha$ emission due to photoionization
by starlight or collisional ionization produced by shocks would be 
strong and readily detectable (I$_{H\alpha}$\,$>$\,10 mR).  
H$\alpha$ emission with an intensity I$_{H\alpha}$\,$\sim$\,100 mR
is observed in many directions toward HVCs that are
known to be located within 10 kpc of the Galactic plane (Tufte, Reynolds,
\& Haffner 1998).
Detectable line emission from gas as far away as the Magellanic 
Stream may be due to  photoionization by photons escaping the Galaxy 
(Bland-Hawthorn \& Maloney 1998)

\subsection{\ovi\ Absorption}
The best diagnostic of hot, collisionally ionized gas at ultraviolet
wavelengths is the \ovi\ doublet at 1031.93, 1037.62 \AA.  The energy 
required to convert O\,{\sc v} into \ovi\ is 114 eV, well above the 
He\,{\sc ii} absorption edge at 54 eV and the ionization potential of 
C\,{\sc iii} at 48 eV.  A plasma
that is radiatively cooling from $\sim$10$^6$ K should have 
N(\ovi)/N(\civ)\,$\sim$\,1--10 (Benjamin \& Shapiro 1999).  Models
for conductive interaces (Borkowski, Balbus, \& Fristrom 1990) and
evolved supernova remnants (Slavin \& Cox 1992; Shelton 1998) yield
similar ratios.  
If the \chvcs\ are extragalactic clouds
photoionized by background radiation, then the amount of \ovi\ in the 
clouds should be roughly 1--2 orders of magnitude less than \civ\ in
the absence of collisional processes. 

We will be able to search for \ovi\ associated with the \chvcs\ with
the Far Ultraviolet Spectroscopic Explorer (see Sembach, this volume).
A column density N(\ovi)\,=\,6.5$\times$10$^{12}$ cm$^{-2}$ will 
have an equivalent width of $\approx$8 m\AA, which is roughly the 
detection limit for S/N\,$\approx$\,30 per FUSE resolution element
for an \ovi\ line with a thermal broadening in 2$\times$10$^{4}$ K
gas (b\,=\,4.6 \kms).
 
 It is interesting to note in this
context that
Cen \& Ostriker (1998) have recently proposed that as much as 50\% of 
the baryons in the nearby universe may be in the form of hot 
(T\,$\sim$\,10$^5$-10$^7$ K) gas.  Thus, a test for the presence of \ovi\
in the \chvcs\ could also place constraints on the amount of gas that
exists in intergalactic clouds at temperatures of 10$^5$-10$^6$ K.

\subsection{HST Data for Other Sight Lines}
Sembach \et (1999) identified two HVCs toward PKS\,2155-304 
($l\,=\,17.7\deg, b\,=\,-52.2\deg$) that appear
to have properties similar to those of the \chvcs\ toward Mrk\,509.
The clouds show \civ\ absorption with little \siii\ absorption.
The velocities are lower than those of the Mrk\,509 \chvcs\ 
($\approx$\,$-$256 and $-140$ \kms), but are consistent with an
extragalactic location.  There is no detectable
21\,cm emission at these velocities directly along the sight line, 
but mapping of the sky nearby 
reveals low column density \hhvcs\ similar to those near the Mrk\,509
sight line.
Further ultraviolet observations of the \chvcs\ toward PKS\,2155-304 
with the HST would allow stronger constraints to be placed on their ionization 
conditions.  
Identification and 
investigation of additional \chvcs\ along other sight lines 
would help constrain the typical kinematical properties 
and covering factors of the clouds, which could then be compared to 
those for higher 
redshift absorption systems.

\end{document}